\documentclass[twocolumn,english,floats,showpacs,prl]{revtex4}
\usepackage[T1]{fontenc}
\usepackage[latin9]{inputenc}
\usepackage{textcomp}
\usepackage{amsmath}
\usepackage{graphicx}
\usepackage{amssymb}

\makeatletter

\DeclareRobustCommand{\greektext}{%
  \fontencoding{LGR}\selectfont\def\encodingdefault{LGR}}
\DeclareRobustCommand{\textgreek}[1]{\leavevmode{\greektext #1}}
\DeclareFontEncoding{LGR}{}{}
\DeclareTextSymbol{\~}{LGR}{126}
\newcommand{\lyxmathsym}[1]{\ifmmode\begingroup\def\b@ld{bold}
  \text{\ifx\math@version\b@ld\bfseries\fi#1}\endgroup\else#1\fi}

\providecommand{\tabularnewline}{\\}

\@ifundefined{textcolor}{}
{%
 \definecolor{BLACK}{gray}{0}
 \definecolor{WHITE}{gray}{1}
 \definecolor{RED}{rgb}{1,0,0}
 \definecolor{GREEN}{rgb}{0,1,0}
 \definecolor{BLUE}{rgb}{0,0,1}
 \definecolor{CYAN}{cmyk}{1,0,0,0}
 \definecolor{MAGENTA}{cmyk}{0,1,0,0}
 \definecolor{YELLOW}{cmyk}{0,0,1,0}
 }

\usepackage[above]{placeins}\usepackage{babel}

\makeatletter

\@ifundefined{textcolor}{}
{
 \definecolor{BLACK}{gray}{0}
 \definecolor{WHITE}{gray}{1}
 \definecolor{RED}{rgb}{1,0,0}
 \definecolor{GREEN}{rgb}{0,1,0}
 \definecolor{BLUE}{rgb}{0,0,1}
 \definecolor{CYAN}{cmyk}{1,0,0,0}
 \definecolor{MAGENTA}{cmyk}{0,1,0,0}
 \definecolor{YELLOW}{cmyk}{0,0,1,0}
 }
\makeatother

\makeatother

\usepackage{babel}

\makeatother

\usepackage{babel}

\makeatother

\usepackage{babel}

\makeatother

\usepackage{babel}

\makeatother

\usepackage{babel}

\begin{document}

\title{Does the isotope effect of mercury support the BCS theory?}

\author{X. Q. Huang$^{1,2}$}

\email{xqhuang@netra.nju.edu.cn}

\affiliation{$^{1}$Department of Physics and National Laboratory of Solid State
Microstructure, Nanjing University, Nanjing 210093, China \\
 $^{2}$ Department of Telecommunications Engineering ICE, PLAUST,
Nanjing 210016, China}

\date{\today}
\begin{abstract}
In this paper, we reexamine the results of isotope effect experiments
of the conventional monoatomic superconductor (Hg). It is shown clearly
that the isotopic coefficients of mercury can be largely deviated
from $\alpha=0.5$, the standard value suggested by the phonon-mediated
BCS pairing theory. According to the reported experimental results
of various mercury isotopes, a giant isotope effect ($\alpha=2.896$)
is numerically found in the data. This study indicates that the validity
of the conventional BCS theory cannot be verified by the isotope effect.
\end{abstract}


\maketitle
In the field of superconductivity, the isotope effect has been considered
can play the role in unraveling the microscopic superconducting mechanism.
Historically, the discovery of the isotope effect in mercury in 1950
was immediately raised to the question regarding the coupling of the
electrons to the lattice vibrations in the superconductor\cite{Maxwell1950,Reynolds1950}.
It was widely believed that the isotope effect may relate to the origin
of the effective attractive interaction between the repulsive electrons,
which leads to the occurrence of superconductivity as suggested by
the BCS pairing theory\cite{BCS1957}. In the framework of BCS pairing
theory, it was predicted that there would be a universal isotopic
coefficient $a=0.5$ for any superconductors\cite{BCS1957}. The condensed
matter physicists concluded that this prediction of the BCS theory
was in good agreement with the reported isotope effects in some conventional
superconductors (e.g., Hg, Sn and Pb)\cite{isotope}.

The experimental facts show that the isotope effect of the vast majority
of superconductors (both conventional and non conventional) largely
deviates from the standard value 0.5 of BCS theory\cite{IV_ie,U_ie,Or_ie,YBCO_ie,Large_ie}.
For the conventional monoatomic superconductors, one notes that some
superconductors have a negligible coefficient $\alpha_{Zr}\simeq0$
, while there are some reports on the inverse isotopic coefficients\cite{IV_ie},
for example, $\alpha_{U}=-2$ for the uranium element\cite{U_ie}.
The inverse isotope effect has also been observed in numerous organic
superconductors\cite{Or_ie}. For the high-temperature superconductors,
the isotope effect can have coefficients both smaller and larger than
0.5 depending on the doping levels. One notes that the cuprate superconductor
$YBa_{2}Cu_{3}O_{7\lyxmathsym{\textminus}y}$ has a very small oxygen
isotope effect $\alpha=0.0\pm0.027$ under the temperature of about
90 Kelvin\cite{YBCO_ie}. Recently, the so-called large iron isotope
effect was reported in the newly discovered iron-based $SmFeAsO_{1-x}F_{x}$
and $Ba_{1-x}K_{x}Fe_{2}As_{2}$ superconductors\cite{Large_ie}.

In this letter, we will argue for the first time that the well-accepted
isotopic coefficient of mercury in fact is not equal to 1/2. It will
be shown that the maximum isotopic coefficient of mercury can reach
as high as 2.896, a value that is about six times of the value predicted
by BCS theory and widely reported in the related papers. This result
indicates that the relationship between the critical temperature $T_{c}$
the isotopic mass $M$ is much more complicated than that of the BCS
theory. It seems most likely that the value of the isotopic coefficient
does not directly lead to any priori conclusion about the pairing
mechanism of the superconductivity. In other words, the BCS theory
of the electron-phonon interaction mechanism not only cannot explain
the isotope effect of the non-conventional superconductors, in fact,
it also cannot explain the isotope effect observed in the conventional
systems.

In the framework of BCS pairing theory, the superconducting critical
temperature $T_{c}$ is given by \begin{equation}
k_{B}T_{c}=1.13\hslash\omega_{D}exp\left(-\frac{1}{VN(E_{F})}\right),\label{tc}\end{equation}
 where $\lyxmathsym{\textgreek{w}}_{D}$ is the Debye frequency, $V$
is the electron\textendash{}phonon interaction strength and $N(E_{F})$
is the electronic density of states at the Fermi surface.

Eq. (\ref{tc}) is considered as one of the most significant and influential
predictions of the BCS theory. In the harmonic approximation, both
$V$ and $N(E_{F})$ are independent of the ionic mass, while the
characteristic frequency $\lyxmathsym{\textgreek{w}}_{D}$ can be
expressed as \begin{equation}
\lyxmathsym{\textgreek{w}}_{D}\propto\frac{1}{\sqrt{M}},\label{db}\end{equation}
 where $M$ is the ionic mass.

By the combination of Eq. (\ref{tc}) and Eq. (\ref{db}), it is not
difficult to conclude that if the product $VN(E_{F})$ is increased,
then the $T_{c}$ will rise. Moreover, using light elements $M$ can
raise the Debye frequency$\lyxmathsym{\textgreek{w}}_{D}$, and this
in turn enhances the $T_{c}$ of the corresponding superconductors.
Based on the above two equations, BCS predicted a maximum $T_{c}$
of around 30K for any superconductors which has been proven wrong
since the discovery of the high-$T_{c}$ oxide superconductors by
Bednorz and M\"uler in 1986\cite{Bednorz1986}.

Within the framework of the electron\textendash{}phonon mechanism,
the $T_{c}$ can be described by the following relation: \begin{equation}
T_{c}=AM^{-\alpha}\label{tc1}\end{equation}
 where $A$ is a constant, $M$ is the mass of the element substituted
by its isotope and $\alpha$ is the so-called isotope coefficient
which is defined as \begin{equation}
\alpha=-\frac{\partial\ln T_{c}}{\partial\ln M}\simeq-\frac{M}{T_{c}}\frac{\Delta T_{c}}{\Delta M},\label{da1}\end{equation}
 where $\Delta T_{c}$ is the shift of the critical temperature and
$\Delta M$ is the difference between the two isotopic mass.

\begin{figure}
\begin{centering}
\resizebox{1\columnwidth}{!}{ \includegraphics{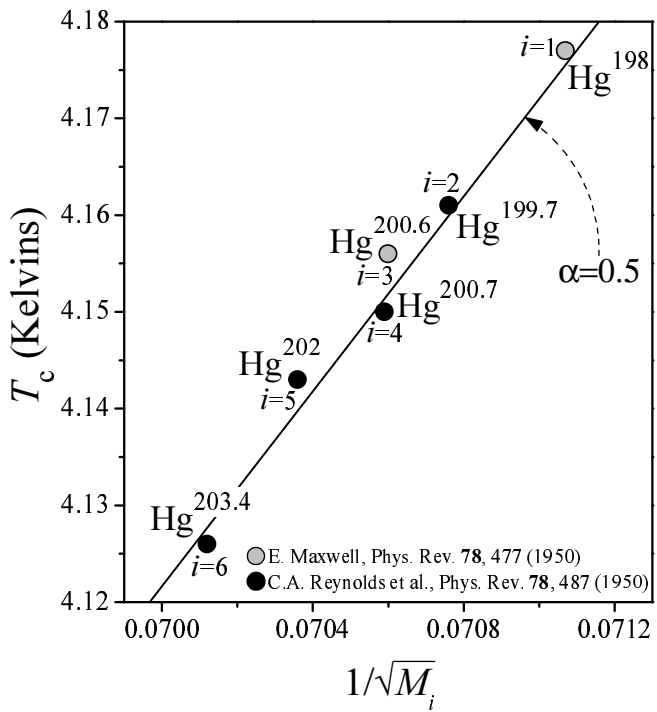}}
\par\end{centering}

\caption{The experimental results of superconducting transition temperature
as a function of isotopic mass for the elemental mercury. }

\label{fig}
\end{figure}

In the standard BCS theory, $T_{c}$ is inversely proportional to
the square root of the masses of the isotopic elements, hence the
isotope-effect coefficient $\alpha=0.5$ which has been considered
in good agreement with the experimental results in some non-transition
metal superconductors, such as Hg, Sn and Pb. In the following discussion,
we will show that this well-known conclusion is evidently wrong.

The superconductivity was first observed in mercury by Onnes in 1911\cite{Onnes}.
Later, the isotope effects for mercury had been intensively investigated
and its isotope-effect was claimed to be around 1/2, as shown in Fig.
\ref{fig}. These results were the basis and foundation of the electron-phonon
mechanism of BCS theory. And now, in order to illustrate the validity
of the BCS theory, the figure has been widely adopted in the textbooks
of superconductivity.

Now let us begin our study of the relationship between the isotope
effect and the BCS theory. For the sake of convenience of discussion
in this letter, the isotopic masses $M_{i}$ and the corresponding
superconducting transition temperatures $T_{c}(i)$ are listed in
the table below. It became obvious that only for the purpose of demonstrating
the correctness of the prediction $\alpha=0.5$ of BCS theory, the
researchers intentionally selected special isotopic samples (1 and
6) and the isotope-effect coefficient was calculated by\begin{equation}
\alpha=-\frac{M_{1}}{T_{c}(1)}\left[\frac{T_{c}(6)-T_{c}(1)}{M_{6}-M_{1}}\right]\approx0.448.\label{da2}\end{equation}

\begin{table}
\begin{ruledtabular} \caption{\label{table-1} The reported superconducting transition temperature
$T_{c}(i)$ of mercury with an isotopic mass of $M_{i}$. Moreover,
$\alpha_{i}$ and $\beta_{i}$ are newly defined parameters of isotope-effect
coefficient and the percent error, respectively. }

\begin{tabular}{c|cccccc}
$Hg(i)$  & $1$  & $2$  & 3  & 4  & 5  & $6$ \tabularnewline
\hline
$M_{i}$  & 198.0  & 199.7  & 200.6  & 200.7  & 202.0  & 203.4\tabularnewline
$T_{c}(i)$  & 4.177  & 4.161  & 4.156  & 4.150  & 4.143  & 4.126\tabularnewline
$\alpha_{i}$  & 0.446  & 0.267  & 2.896  & 0.260  & 0.592  & \tabularnewline
$\beta_{i}$  & 10.8  & 46.6  & 479.2  & 48.0  & 18.4  & \tabularnewline
\end{tabular}\end{ruledtabular}
\end{table}

With the two sets of data ($T_{c}(1)$, $M_{1}$) and ($T_{c}(6)$,
$M_{6}$), one indeed can obtain the desired isotope value close to
1/2, but this is completely wrong. Mathematically, to apply Eq. (\ref{da1})
to estimate the parameter $\alpha$, the variable quantities $\Delta T_{c}$
and $\Delta M$ must be infinitesimal. Hence, the calculation formula
(\ref{da2}) holds only when $\Delta T_{c}$ and $\Delta M$ satisfy
a linear relationship which is obviously not consistent with Eq. (\ref{tc1}).
Here, it is argued that a more accurate formula for calculating the
isotope-effect coefficient would be \begin{eqnarray}
\alpha_{i} & \approx & -\frac{\ln T_{c}(i+1)-\ln T_{c}(i)}{\ln M_{i+1}-\ln M_{i}}\notag\label{da3}\\
 & \approx & -\frac{M_{i}}{T_{c}(i)}\frac{T_{c}(i+1)-T_{c}(i)}{M_{i+1}-M_{i}}.\end{eqnarray}

In the above formula, two sets of adjacent data ($T_{c}(i)$, $M_{i}$)
and ($T_{c}(i+1)$, $M_{i+1}$) are applied in the numerical simulation.
We have argued that for a given superconductor, its isotope-effect
coefficient may be completely different. This argument is well confirmed
by directly putting the reported experimental results of Fig. \ref{fig}
into Eq. (\ref{da3}), as shown in Table \ref{table-1}. It is not
difficult to find that all the obtained $\alpha_{i}$ are very different
from 0.5 (in the range from 0.2 to 3), surprisingly, a giant isotope
effect ($\alpha=2.896$ ) which is about six times higher than the
value 1/2 suggested by BCS theory. In order to better present the
deviation of the experimental results from the BCS theory, we define
the percent error $\beta_{i}$ as: \begin{equation}
\beta_{i}=\frac{\left|\alpha_{i}-0.5\right|}{0.5}\times100\%.\end{equation}
 where 0.5 is the full isotope effect in the framework of BCS theory.

As also shown in Table \ref{table-1}, the minimum percent error of
the isotope coefficient for the elemental mercury is about 10.8\%,
while the maximum percent error can reach an inconceivable value of
479.2\%. These results imply that the validity of the full isotope
effect ($\alpha=0.5$) of BCS pairing theory has never been experimentally
verified. It is a possibility that the electron-phonon interaction
based BCS theory may be fundamentally flawed.

In this short letter, we have presented arguments against the mainstream
view that the isotope effect in the non-transition metal superconductor
(Hg) is equal to 1/2, which is the characteristic value predicted
by the classical form of BCS theory. It has been pointed out that
the traditional method of calculation used to estimate superconducting
isotope effect is mathematically unreliable. Based on the reported
experimental result, very different isotope coefficients ranged from
0.267 to 2.896 for the mercury have been numerically obtained by using
the improved method. Our results imply that the dependence of the
critical temperature for superconductivity upon the isotopic mass
was much more complicated than the BCS theory previously advocated.
In our opinion, the isotope effect cannot be applied as the direct
evidence of the proposed electron\textendash{}phonon coupling mechanism.


\begin{thebibliography}{12}
\bibitem{Maxwell1950}E. Maxwell, Phys. Rev. \textbf{78}, 477 (1950).

\bibitem{Reynolds1950}C. A. Reynolds, B. Serin, W. H. Wright, and
L. B. Nesbitt, Phys. Rev. \textbf{78}, 487 (1950).

\bibitem{BCS1957}J. Bardeen, L. N. Cooper, and J. R. Schrieffer,
Phys. Rev. \textbf{108}, 1175 (1957).

\bibitem{isotope}http://hyperphysics.phy-astr.gsu.edu/hbase/ solids/coop.html\#c5

\bibitem{Crawford1990}M. K. Crawford, W. E. Farneth, E. M. McCarronn
III, R. L. Harlow and A. H. Moudden, Science \textbf{250,} 1390 (1990).

\bibitem{IV_ie} P. M. Shirage \emph{et al}., Phys. Rev. Lett. \textbf{103},
257003 (2009).

\bibitem{Large_ie}R. H. Liu, T. Wu, G. Wu, H. Chen, X. F. Wang, Y.
L. Xie, J. J. Yin, Y. J. Yan, Q. J. Li, B. C. Shi, W. S. Chu, Z. Y.
Wu, X. H. Chen, Nature \textbf{459}, 64 (2009).

\bibitem{U_ie}R. D. Fowler, J. D. G. Lindsay, R. W. White, H. H.
Hill, and B. T. Matthias, Phys. Rev. Lett. \textbf{19}, 892 (1967).

\bibitem{YBCO_ie}L. C. Bourne \emph{et al.}, Phys. Rev. Lett. \textbf{58}
, 2337 (1987).

\bibitem{Or_ie}J. A. Schlueter \emph{et al}., Physica C \textbf{265},
163 (1996).

\bibitem{Bednorz1986} J. G. Bednorz and K. A. M\"uler, Z. Phys. B \textbf{64},
189 (1986).

\bibitem{Onnes}H. K. Onnes, Comm. Phys. Lab. Univ. Leiden, Nos. 122
and 124 (1911).
\end{thebibliography}
\end{document}